\documentclass[a4paper, 11pt]{article}

\usepackage{a4wide}
\usepackage{latexsym}
\usepackage{graphicx}
\usepackage{color}
\usepackage[hypertex]{hyperref}
\usepackage{booktabs}
\usepackage{apacite}

\title{A bibliometric index based on the complete list of cited publications}

\author{
Mark Levene and Trevor Fenner \\
Department of Computer Science and Information Systems \\
Birkbeck, University of London \\
London, WC1E 7HX, U.K. \\ \{mark,trevor\}@dcs.bbk.ac.uk \\
\and Judit Bar-Ilan  \\ Department of
Information Science \\ Bar-Ilan University \\ 52900 Ramat Gan, Israel \\
barilaj@mail.biu.ac.il }

\date{}

\begin{document}

\maketitle

\newtheorem{theorem}{Theorem}[section]
\newtheorem{corollary}[theorem]{Corollary}
\newtheorem{lemma}[theorem]{Lemma}
\newtheorem{proposition}[theorem]{Proposition}
\newtheorem{definition}{Definition}[section]
\newtheorem{algorithm}{Algorithm}
\newtheorem{example}{Example}[section]

\begin{abstract}

We propose a new index, the $j$-index, which is defined for an author as the sum of the square roots of the numbers of citations to each of the author's publications. The idea behind the $j$-index it to remedy a drawback of the $h$-index $-$ that the $h$-index does not take into account the full citation record of a researcher. The square root function is motivated by our desire to avoid the possible bias that may occur with a simple sum when an author has several very highly cited papers. We compare the $j$-index to the $h$-index, the $g$-index and the total citation count for three subject areas using several association measures.

Our results indicate that that the association between the $j$-index and the other indices varies according to the subject area. One explanation of this variation may be due to the proportion of citations to publications of the researcher that are in the $h$-core. The $j$-index is {\em not} an $h$-index variant, and as such is intended to complement rather than necessarily replace the $h$-index and other bibliometric indicators, thus providing a more complete picture of a researcher's achievements.

\end{abstract}


\section{Introduction}

In a broad sense, the number of publications of a researcher is a measure of quantity and the total number of citations to these publications is often perceived as a measure of quality.
Although these metrics each take into account only one facet of a researcher's impact, several other bibliometric indices, such as the $h$-index and the $g$-index, combine citation and publication counts.
However, the $h$-index and its derivatives \cite{BORN11} have the drawback that they do not take into account the full citation list of a researcher, but, on the other hand, the total citation count has the drawback of biasing the index in favour of researchers with very highly cited top papers or very many papers with a relatively small number of citations. We first briefly review the $h$-index and some of its variants, and then introduce the {\em j-index}, a new index that addresses some of the drawbacks mentioned.

\medskip

The {\em h-index} of a researcher is the maximum number $h$ of the researcher's publications that each have at least $h$ citations \cite{HIRS05}. As an equivalent definition, rank a researcher's publication list in descending order of the number of citations, with paper $i$ receiving $C(i)$ citations. The $h$-index is then the largest rank $h$ for which $C(h) \ge h$. The $h$-index is completely insensitive to the fact that a researcher's top few papers are very highly cited, and conversely also to a researcher having many papers with very few citations \cite{BORN07b}. A suggested improvement over the $h$-index, which gives more weight to highly cited papers, is the {\em g-index}. The $g$-index of a researcher is the largest rank $g$ for which $\sum_{i=1}^g C(i) \ge g^2$ \cite{EGGH06}; it is easily shown that $g \ge h$.
A problem with the $g$-index is that it may still be biased since, if a researcher has a few papers that are very highly cited and the rest have very few citations, the $g$-index will still be high. This is because the $g$-index is equal to the largest rank $g$ such that the average number of citations up until that rank is at least $g$. (Note that we consider the variant of the $g$-index  that is not limited by the actual number of publications, i.e. fictitious papers with zero citations may be added to satisfy the definition of the $g$-index \cite{EGGH06}.)
If the $h$-index of a researcher is $h$, the {\em h-core} is the set of her/his $h$ most highly cited publications.
(It is irrelevant which of the publications with exactly $h$ citations are chosen.)
Another attempt to try and address the fact that the $h$-index does not take into account the total number of citations to papers in the $h$-core is via the {\em A-index}, which is the average number of citations to papers in the $h$-core, i.e. $A = 1/h \sum_{i=1}^h C(i)$  \cite{BIHU07}. However, the $A$-index suffers from the fact that taking an average will, all other things being equal, often favour authors with fewer publications when they are highly cited. To remedy this, the {\em R-index} has been proposed, where $R = \sqrt{\sum_{i=1}^h C(i)} = \sqrt{A h}$ \cite{BIHU07}. It is easy to see that $h \le R \le A$. However, the $A$ and $R$ indices, and to a lesser extent the $g$-index, ignore the effect of papers outside the $h$-core, which are also part of a researcher's output.

\medskip

In this paper we take a step towards defining an index that takes into account both the quantity and the quality of a researcher's
output as reflected by the citation data, but that does not suffer from the drawbacks of simply counting the number of citations.
On the one hand, the issue of quantity is addressed by an index which considers all of a researcher's cited papers, so that each cited output contributes towards the index. On the other hand, the issue of quality is addressed by applying a function to the numbers of citations that has the effect of reducing the impact of very highly cited papers, which tends to bias the values of many bibliometric indices. As an example, consider a researcher $\alpha$ who published a single paper with 100 citations compared to a researcher $\beta$ who published 10 papers, each having 10 citations. The total number of citations of $\alpha$ and $\beta$ are the same, but the average number of citations strongly favours $\alpha$, who has far fewer publications. The $h$-index resolves this particular issue by strongly favouring $\beta$. However, if researcher $\alpha$ publishes another 9 papers each having 10 citations, their $h$ indices will be same, although it is now obvious that researcher $\alpha$ has had more impact on the field. The $R$-index addresses this problem but ignores publications outside the $h$-core: if researcher $\beta$ also has a long string of publications each with fewer than 10 citations, these will have no effect on $R$, $A$ or any other index that only takes into account publications in the $h$-core.

\medskip

Here we propose a new index, called the {\em j-index}, that takes into account all cited publications.
By doing so, it is fairer to researchers who have a long tail of publications outside the $h$-core (see Section~\ref{sec:analysis}). Moreover, the $j$-index also reduces the bias that some indices tend to introduce in favour of researchers having a small number of very highly cited papers.

The function we use for defining the $j$-index is the the sum of the square roots of the numbers of citations.
This function arises in the study of {\em social welfare functions} \cite{SEGA06}.
In that context, maximising the sum of the utilities (in our case, the total number of citations) is the utilitarian solution, where a ``good'' is allocated to the individual with the highest utility, while maximising the sum of the square roots is the optimal solution when randomisation, with probabilities proportional to individual utilities, is used to decide to which individual the ``good'' is allocated.

\smallskip

In our context of a bibliometric indicator, the sum of square roots serves to dampen the effect of highly cited papers, yet take into account the full citation list. Thus we propose the {\em j-index} of the researcher, defined as $j = \sum_{i=1}^n \sqrt{C(i)}$, where $n$ is the number of cited publications. Although we do not claim the $j$-index is optimal in the sense that it is for welfare functions, we will demonstrate that it addresses some of the problems associated with the $h$-index and its variants.

\smallskip

We note that we do {\em not} address variants of indices that may arise from taking into account self-citations, multi-author publications, field dependence, and the age of publications \cite{BORN11}. Although such refinements to indices are obviously worth pursuing, it would bias our comparison the total citation count and the $h$ and $g$ indices, which, in their original forms, do no not take such potential improvements into account.

\medskip

The rest of the paper is organised as follows.
In Section~\ref{sec:j-index} we introduce the $j$-index and also an appropriate smoothing operator, and
in Section~\ref{sec:comparison} we compare the $j$-index with the $h$-index and other bibliometric indicators in the context of data sets from three subject areas taken from ISI's Highly Cited Database.
In Section~\ref{sec:manipulate} we demonstrate that the $j$-index cannot be easily manipulated by adding publications with single citations, which may
actually turn out to be self-citations, and
in Section~\ref{sec:analysis} we analyse the $h$-index in terms of the proportion of citations to publications inside and outside the $h$-core.
Finally, in Section~\ref{sec:conc} we give our concluding remarks.

\section{The j-index}
\label{sec:j-index}

We assume that a researcher's publication list is ranked in descending order of the number of citations, with paper $i$ receiving $C(i)$ citations, and that $n$ is the number of cited publications.

We define {\em j-index} as
\begin{equation}\label{eq:j-index}
j = \sum_{i=1}^n \sqrt{C(i)}.
\end{equation}

We observe that the $j$-index is a sum of square roots, whereas the $R$-index is the square root of a sum, so clearly $R \le j$. Moreover, the $j$-index always takes into account the full range of cited publications, unlike the previously mentioned indices that are restricted to the $h$-core, or the $g$-index. We thus stress that the $j$-index is not an $h$-index variant.

\smallskip

In addition, we define a smoothing operator ${\mathcal S}$ for a monotonically decreasing sequence $u(i)$, defined such that
\begin{equation}\label{eq:j-values}
{\mathcal S}u(i) = \frac{1}{i} \sum_{k=1}^i u(k).
\end{equation}

\noindent
Clearly the ${\mathcal S}$ operator maintains monotonicity, i.e. ${\mathcal S}u(p) \ge {\mathcal S}u(q)$ if  $p \le q$.

\smallskip

We now define the $j {\mathcal S}${\em -index} as in (\ref{eq:j-index}) but using the smoothed values ${\mathcal S}C(i)$ rather than the raw values $C(i)$.
Using the smoothed values is similar to the computation of a moving average for a time series \cite{CHAT96}. We note that ${\mathcal S}C(1)$ is the maximum number of citations and ${\mathcal S}C(n)$ is the average number.

\section{Comparing the j-index with the h-index}
\label{sec:comparison}

We compare the $j$-index with the $h$-index and, in order to get a more comprehensive picture, we also include the $g$-index and the total citation count $T$ in the comparison. According to \cite{BIHU07}, the $h$, $g$, $A$ and $R$ indices are highly correlated, which is why we did not also include the $A$ and $R$ indices in the comparison.

Our comparison is based on comparing two lists of rankings of researchers using
three well-understood association measures: the Spearman correlation coefficient \cite{MOTU95}, the Spearman footrule \cite{DIAC77}
and the $M$-measure \cite{BARI07}.

\smallskip

Suppose that we are ranking $n$ researchers, labelled $1,2,\ldots,n$, according to two criteria, and that $\sigma_1(i)$ and $\sigma_2(i)$ are the rankings of the $i$th researcher according to the first and second critera, respectively. The Spearman rank correlation coefficient is given by
\begin{displaymath}
1 - \frac{6 \sum_{i=1}^n \left(\sigma_1(i) - \sigma_2(i)\right)^2}{n (n^2 -1)}.
\end{displaymath}

Spearman's footrule is a useful alternative measure for comparing the orderings of two permutations; it is given by
\begin{displaymath}
1 - \frac{\sum_{i=1}^n |\sigma_1(i) - \sigma_2(i)|}{maxF},
\end{displaymath}
where $maxF$, the normalisation factor, is chosen so that the minimum value of the measure is zero, and is given by
\begin{displaymath}
maxF = \left \lfloor \frac{n^2}{2} \right \rfloor.
\end{displaymath}

The $M$-measure is a weighted variation of Spearman's footrule, giving more weight to identical or near identical rankings
among the researchers in the top positions. It attempts to capture the intuition that identical or near identical rankings among the top researchers indicates greater similarity between the rankings. It is given by
\begin{displaymath}
1 - \frac{\sum_{i=1}^n \left| \frac{1}{\sigma_1(i)} - \frac{1}{\sigma_2(i)} \right|}{maxM},
\end{displaymath}
where $maxM$, the normalisation factor,  is chosen so that the minimum value of the measure is zero, and is given by
\begin{displaymath}
maxM = \sum_{i=1}^n \left|\frac{1}{i} - \frac{1}{n-i+1}\right|.
\end{displaymath}
\medskip

In the tables below we make use of the following notation to indicate the level of significance of the Spearman rank correlation coefficient:
\begin{description}
\item (**) indicates that a 2-tailed correlation test is significant at the 0.01 level.		

\item (*) indicates that a 2-tailed correlation test is significant at the 0.05 level.		

\item (n) indicate that a 2-tailed correlation test is not significant at the 0.05 level.
\end{description}
\smallskip

For the empirical comparison, we chose three subject areas: Immunology, Economics and Physics, from the medical, social and physical sciences, respectively. For each of the these areas, ISI's Highly Cited Database (\url{www.isihighlycited.com}) was consulted, and 20 names were selected. The names were selected in such a way that the researchers were still active and their publications could be easily disambiguated when there were multiple authors with the same name. The publication lists with the citation counts for these subject areas were downloaded from Thomson-Reuters (ISI) Web of Science at the end of September 2010. (See the Appendix for a summary of the researchers' data sets that we used, together with the various indices we computed. These are sorted in descending order of the $h$-index.)

\smallskip

In Tables \ref{table:corr-immunology}, \ref{table:corr-economics} and \ref{table:corr-physics}, we show the correlation analyses for the Immunology, Economics and Physics researchers, respectively. We compared $T$, $h$ and $g$ with both the $j$ and $j {\mathcal S}$ indices.
The general patterns for Immunology and Economics exhibit high similarity  between these three indices and both $j$ indices. The similarities for Immunology are noticeably lower than for Economics when using the $M$-measure. So, for both  Immunology and Economics, there is strong agreement when ranking the researchers using $T$, $h$ and $g$, on the one hand, and $j$ and $j {\mathcal S}$, on the other. However, the ordering of the top Economics researchers is generally agreed upon, while for Immunology this is not the case (see the $M$-measure values in Tables \ref{table:corr-immunology} and \ref{table:corr-economics}). We further note that total number of citations $T$ has a high correlation with $j$ and $j {\mathcal S}$, implying that for the group of Immunology and Economics researchers $T$ is also a reasonable metric. (We note that using the {\em average} citation count would be unsatisfactory, as it tends to favour researchers who have published fewer papers but whose papers are highly cited vis-a-vis those who have in addition published a number of less frequently cited papers.)

\begin{table}[htbp]
  \centering
  \caption{Correlation analysis for Immunology researchers}
    \begin{tabular}{cccc|ccc}
    \addlinespace
    \toprule
          & \multicolumn{3}{c}{$j$-index} & \multicolumn{3}{c}{$j {\mathcal S}$-index} \\
    \midrule
              & Spearman  & Footrule & $M$   & Spearman  & Footrule & $M$ \\
    $T$       & 0.847(**) &	0.770	 & 0.674 & 0.884(**) & 0.820	& 0.705 \\
    $h$	      & 0.953(**) &	0.870	 & 0.605 & 0.919(**) & 0.840	& 0.619 \\
    $g$	      & 0.765(**) & 0.700	 & 0.533 & 0.806(**) & 0.740	& 0.561 \\
    $j$   	  &	$-$		  & $-$      & $-$   & 0.973(**) & 0.930    & 0.962  \\
    $j {\mathcal S}$       &	0.973(**) &	0.930	 & 0.962 & $-$       & $-$      & $-$   \\			
    \bottomrule
    \end{tabular}
  \label{table:corr-immunology}
\end{table}

\begin{table}[htbp]
  \centering
  \caption{Correlation analysis for Economics researchers}
    \begin{tabular}{cccc|ccc}
    \addlinespace
    \toprule
          & \multicolumn{3}{c}{$j$-index} & \multicolumn{3}{c}{$j {\mathcal S}$-index} \\
    \midrule
              & Spearman & Footrule & $M$ & Spearman  & Footrule & $M$ \\
    $T$       & 0.874(**) & 0.770 & 0.899 & 0.943(**) & 0.830    & 0.888 \\
    $h$         & 0.910(**) & 0.800 & 0.852 & 0.850(**) & 0.750    & 0.821 \\
    $g$         & 0.886(**) & 0.770 & 0.889 & 0.941(**) & 0.830    & 0.877 \\
    $j$         &  $-$      & $-$   & $-$   & 0.962(**) & 0.900    & 0.921 \\
    $j {\mathcal S}$       & 0.962(**)  & 0.900 & 0.921 & $-$       & $-$       & $-$  \\
    \bottomrule
    \end{tabular}
  \label{table:corr-economics}
\end{table}

\begin{table}[htbp]
  \centering
  \caption{Correlation analysis for Physics researchers}
    \begin{tabular}{cccc|ccc}
    \addlinespace
    \toprule
          & \multicolumn{3}{c}{$j$-index} & \multicolumn{3}{c}{$j {\mathcal S}$-index} \\
    \midrule
              & Spearman       & Footrule & $M$   & Spearman       & Footrule & $M$ \\
    $T$       & {\bf 0.441(n)} & 0.470    & 0.286 & 0.764(**)      & 0.670 & 0.457 \\
    $h$         & {\bf 0.332(n)} & 0.400    & 0.184 & {\bf 0.371(n)} & 0.460 & 0.231 \\
    $g$         & {\bf 0.023(n)} & 0.280    & 0.164 & {\bf 0.468(*)} & 0.500 & 0.338 \\
    $j$         & $-$            & $-$      & $-$   & 0.836(**)      & 0.750 & 0.603 \\
    $j {\mathcal S}$            & 0.836(**) & 0.750         & 0.603 & $-$            & $-$ &  $-$ \\
    \bottomrule
    \end{tabular}
  \label{table:corr-physics}
\end{table}

The Physics group appears to be an outlier as the correlations are all lower and less significant. The highest association in this case is between $T$ and $j {\mathcal S}$. To get a better picture, we consider the associations between $T$, $h$ and $g$.
For Immunology and Economics, these are all high and significant at the 0.01 level. Table~\ref{table:full-corr} shows these associations for the Physics researchers. We observe that $g$ is more correlated with $T$ than $h$ is, which is not surprising since $g$ takes into account citations to some publications outside the $h$-core.
We note that there is a stronger association between $g$ and $j {\mathcal S}$ than between $h$ and $j {\mathcal S}$. However, surprisingly, this is the other way around for the $j$-index. This, together with the fact that most of the correlations are higher with
$j {\mathcal S}$ than with $j$ is an indication that it may be preferable to use the smoothed rather than the raw data.

\begin{table}[htbp]\small
  \centering
  \caption{Full correlation analysis for Physics researchers}
    \begin{tabular}{cccc|ccc|ccc}
    \addlinespace
    \toprule
          & \multicolumn{3}{c}{$T$} & \multicolumn{3}{c}{$h$}    & \multicolumn{3}{c}{$g$} \\
    \midrule
              & Spearman        & Footrule & $M$    & Spearman        & Footrule & $M$     & Spearman      & Footrule & $M$ \\
    $T$       & $-$             & $-$      & $-$    & 0.585(**)       & 0.630    & 0.658 & 0.890(**)       & 0.790    & 0.874 \\
    $h$        & 0.585(**)       & 0.630    & 0.658  &  $-$            & $-$      & $-$   & {\bf 0.499(*)}  & 0.570    & 0.665 \\
    $g$         & 0.890(**)       & 0.790    & 0.874  & {\bf 0.499(*)}  & 0.570    & 0.665 & $-$             & $-$      & $-$  \\
    $j$         & {\bf 0.441(n)}  & 0.470    & 0.286  & {\bf 0.332(n)}  & 0.400    & 0.184 & {\bf 0.023(n)}  & 0.280    & 0.164 \\
    $j {\mathcal S}$             & 0.764(**)       & 0.670    & 0.457  & {\bf 0.371(n)}  & 0.460    & 0.231 & {\bf 0.468(*)}  & 0.500    & 0.338 \\
    \bottomrule
    \end{tabular}
  \label{table:full-corr}
\end{table}

One conclusion from the above analysis is that the total citation count is an important index that should be taken into account, since it significantly correlates with the other measures. Still, a word of caution is appropriate here: $T$ is biased by the highly cited papers, which is one of the issues addressed by the $h$-index and its variants. Our justification for the $j$-index is that it tries to resolve this issue with $T$, while at the same time addressing some of the problems with the $h$ and $g$ indices, which are, respectively, unaffected and less affected, by the lower-cited publications.

\section{Manipulating the j-index}
\label{sec:manipulate}

One possible argument against $j$-index may be that it can be manipulated by an author with many publications each having a single citation. Taking this argument further one may even assume that these single citations are self-citations. In order to investigate this possibility we carried out a further analysis on our data set by first removing all publications with a single citation, and in a second analysis decreasing all citations to publications by one.

In all three disciplines the new data sets show very little relative change in the rankings according to the $j$-index.
More specifically, for Immunology there was also a single change, between Aarden and Goodnow in positions 14 and 15, in Economics there was a single change (only  for the case when publications with single citations were removed) between Reinganum and Galor at positions 18 and 19, and
for Physics there was a single change Alivisatos and Foxon in positions 7 and 8.

As the correlations were computed on the relative rankings of researchers, we can conclude the $j$-index is not very sensitive to small changes in the citation patterns, and thus cannot be easily manipulated by adding papers with single self-citations.
If we relax the constraint of a single self-citation, then the problem will be exacerbated, but we note for the $j$-index, self citing lower ranked papers will have a greater proportional effect. Therefore, to tackle this problem it may be useful to completely ignore the citations in the tail when computing the $j$-index, although further research has to be carried out to substantiate this.

\smallskip

In comparison the $h$-index will only change by at most one in these cases. However, as was shown in \cite{BART11}, in the more general case, authors can strategically (rather than randomly) self-cite their papers and considerably inflate their $h$-index.

\section{Analysis of the h-index}
\label{sec:analysis}

We now analyse the $h$-index by partitioning the publications according to whether or not they are present in the $h$-core. Our contention is that the $h$-index is less satisfactory when there are a significant number of citations to publications outside the $h$-core.

\smallskip

Recall that $T$ is the total number of citations for a researcher.
We define $H_1$ to be the number of citations to publications in the $h$-core, i.e. $H_1 = Ah = R^2$ (recalling the definitions of the $A$ and $R$ indices from the introduction), $H_2$ to be the minimum possible number of citations to publications in the $h$-core, i.e. $H_2 = h^2$, $H_3$ to be the number of ``excess citations'' to publications in the $h$-core, i.e. $H_3 = H_1 - H_2$, and $H_4$ to be the number of citations to publications outside the $h$-core, i.e. $H_4 = T-H_1$. Note that $H_1 + H_4 = H_2 + H_3 + H_4 = T$. We now define $G_i = H_i/T$, i.e. the proportion of citations corresponding to $H_i$.
In Table~\ref{table:g} we show the averages of these numbers for the three data sets.


\begin{table}[ht]
\begin{center}
\begin{tabular}{|l|c|c|c|c|c|c|c|c|}\hline
Discipline & $H_1$      & $H_2$   & $H_3$  & $H_4$   & $G_1$ & $G_2$  & $G_3$ & $G_4$ \\ \hline
Immunology & 15860.05 &	4888.45	& 10971.60 & 4010.65 & 0.798 & 0.246  &	0.552 &	0.202 \\
Economics  & 5100.85  &	929.605	& 4171.25  & 431.25	 & 0.922 & 0.168  & 0.754 & 0.078 \\
Physics    & 14947.50 & 4205.30	& 10742.20 & 5983.80 & 0.714 & 0.201  & 0.513 & 0.286 \\ \hline
\end{tabular}
\end{center}
\caption{\label{table:g} Average proportions of citations inside and outside the $h$-core}
\end{table}
\smallskip

For all three data sets, a significant majority of the citations are to publications in the $h$-core, which is due to the long-tailed distribution of citations (cf. \cite{REDN98}). In this respect Economics stands out, with $G_4 = 0.078$ indicating that there are very few citations to publications outside the $h$-core. Moreover, looking at the $G_3$ values, we see that around three quarters of the citations to the $h$-core are ``excess citations'', which is essentially the main motivation for the $g$-index. Now, looking at $G_4$, i.e. the citations to publications outside the $h$-core, we observe that for Physics and Immunology, with $G_4 = 0.286$ and $G_4 = 0.202$, respectively, these represent a significant proportion of the citations. This is a clear indication that it is not sufficient to consider only the $h$-core, rather the complete citation and publication patterns of the researchers should be taken into account.
The suggested $j$-index is a step towards achieving this goal. We stress that we do not suggest simply replacing the existing indices with the new index, but rather that it should be used to supplement them in order to provide a more complete picture of the scientists' achievements.

\section{Concluding Remarks}
\label{sec:conc}

We propose a new bibliometric measure, the $j$-index, that takes all of the citations to a researcher's publications into account. The $j$-index thus complements the $h$-index rather than being a variant of it. We used data sets of researchers from three areas, Immunology, Economics and Physics, and we have have compared the difference in the rankings by the $j$-index with those by the $h$-index, $g$-index and total citation count. The association between the rankings is highest for the Economics group. It is not quite so high for the Immunology group, and is much lower for the Physics group. The varying association can be partly explained by the differing average proportions of citations to publications outside the $h$-core for the three groups. We suggest that the $j$-index may be particularly useful for subject areas where this proportion is significant.

The smoothing of the $j$-index by using the $j {\mathcal S}$-index was also proposed, and this generally has the effect of increasing the associations with the other bibliometric indices. However, more research needs to be done on the effect of using smoothed values rather than raw ones in computing the $j$-index, and whether there may also be advantages in computing the other indices using smoothed values.

\newpage
\section{Appendix: Researchers Data Sets}
\label{sec:appendix}

\begin{table}[h]
  \centering
  \caption{Immunology researchers data}
    \begin{tabular}{lrrrrrrrr}
    \addlinespace
    \toprule
    Name  & \#pub & \#cited & $T$     & $h$     & $g$     & $j$-index & $j {\mathcal S}$-index & $G_1$ \\
    \midrule
    Marrack, Philippa C. & 445   & 326   & 45130 & 103   & 208   & 3048.9 & 5865.1 & 0.812 \\
    Nadler, Lee Marshall & 468   & 312   & 32422 & 101   & 174   & 2569.5 & 4743.1 & 0.792 \\
    Gleich, Gerald J. & 891   & 659   & 35065 & 96    & 164   & 3745.3 & 7388.1 & 0.619 \\
    Janossy, George  & 490   & 384   & 26430 & 93    & 148   & 2610.0 & 4694.5 & 0.688 \\
    Shevach, Ethan M. & 472   & 341   & 33000 & 93    & 175   & 2638.2 & 5138.5 & 0.781 \\
    Ravetch, Jeffrey V. & 186   & 165   & 22743 & 78    & 150   & 1614.0 & 2843.8 & 0.874 \\
    Krieg, Arthur M. & 308   & 232   & 21413 & 72    & 143   & 1756.3 & 3430.9 & 0.810 \\
    Figdor, Carl Gustav & 328   & 262   & 17665 & 69    & 126   & 1698.5 & 3328.8 & 0.769 \\
    Takeuchi, Osamu  & 185   & 166   & 24307 & 68    & 157   & 1462.6 & 3194.7 & 0.926 \\
    Hamaoka, Toshiyuki  & 498   & 424   & 15717 & 64    & 106   & 2096.0 & 3947.3 & 0.590 \\
    Goodnow, Christopher C. & 199   & 152   & 13723 & 60    & 116   & 1154.8 & 2172.9 & 0.874 \\
    Kehrl, John H. & 238   & 156   & 14897 & 59    & 121   & 1180.9 & 2379.8 & 0.850 \\
    Adorini, Luciano  & 289   & 238   & 11966 & 58    & 100   & 1390.1 & 2516.7 & 0.698 \\
    Aarden, Lucien A. & 215   & 169   & 12854 & 57    & 111   & 1157.6 & 2259.3 & 0.840 \\
    Delespesse, Guy  & 294   & 207   & 9132  & 57    & 87    & 1154.2 & 1972.9 & 0.701 \\
    Bendelac, Albert  & 139   & 115   & 11983 & 52    & 109   & 951.7 & 1720.8 & 0.890 \\
    Malefyt, Rene DeWaal & 129   & 97    & 18561 & 50    & 136   & 1013.1 & 2086.3 & 0.943 \\
    Bjorkman, Pamela J. & 174   & 141   & 15969 & 46    & 126   & 984.4 & 2496.3 & 0.908 \\
    Parronchi, Paola & 100   & 86    & 8159  & 36    & 90    & 621.8 & 1305.8 & 0.918 \\
    Samraoui, Boudjema  & 28    & 19    & 6278  & 8     & 80    & 179.0 & 549.4 & 0.994 \\
    \bottomrule
    \end{tabular}%
  \label{tab:addlabel1}%
\end{table}%

\begin{table}[h]
  \centering
  \caption{Economics researchers data}
    \begin{tabular}{lrrrrrrrr}
    \addlinespace
    \toprule
    Name  & \#pub & \#cited & $T$     & $h$     & $g$     & $j$-index & $j {\mathcal S}$-index & $G_1$ \\
    \midrule
    Kahneman, Daniel  & 122   & 110   & 35162 & 58    & 188   & 1373.0 & 3197.0 & 0.962 \\
    Stiglitz, Joseph E. & 214   & 188   & 14654 & 55    & 118   & 1235.9 & 2650.9 & 0.832 \\
    Diebold, Francis X. & 87    & 76    & 4075  & 36    & 63    & 439.1 & 835.0 & 0.910 \\
    Milgrom, Paul Robert & 48    & 47    & 9168  & 35    & 95    & 553.8 & 922.3 & 0.981 \\
    Maskin, Eric  & 72    & 63    & 4163  & 33    & 64    & 422.8 & 748.0 & 0.915 \\
    Zajac, Edward J. & 49    & 44    & 3964  & 31    & 62    & 354.5 & 601.8 & 0.953 \\
    Lakonishok, Josef  & 57    & 51    & 3229  & 29    & 57    & 345.3 & 576.3 & 0.919 \\
    Besley, Timothy J. & 89    & 67    & 2388  & 28    & 48    & 336.0 & 572.3 & 0.854 \\
    Hendry, David F. & 129   & 112   & 4282  & 28    & 64    & 503.3 & 1106.7 & 0.846 \\
    Oswald, Andrew J. & 79    & 67    & 3063  & 27    & 55    & 362.1 & 671.1 & 0.871 \\
    Akerlof, George A. & 56    & 49    & 5670  & 26    & 75    & 373.6 & 842.6 & 0.958 \\
    Rodrik, Dani  & 84    & 67    & 2963  & 25    & 53    & 357.4 & 673.7 & 0.849 \\
    Caballero, Ricardo J. & 63    & 59    & 1848  & 23    & 42    & 286.6 & 458.4 & 0.807 \\
    Rotemberg, Julio J. & 59    & 49    & 2259  & 23    & 47    & 272.1 & 489.6 & 0.886 \\
    Bernanke, Ben S. & 45    & 38    & 2525  & 20    & 50    & 237.8 & 463.0 & 0.954 \\
    Constantinides, George M. & 40    & 31    & 1895  & 20    & 43    & 204.2 & 343.4 & 0.957 \\
    Gali, Jordi  & 37    & 35    & 3110  & 19    & 55    & 248.8 & 500.3 & 0.959 \\
    Galor, Oded  & 39    & 38    & 2067  & 19    & 45    & 221.2 & 418.5 & 0.935 \\
    Reinganum, Jennifer F. & 46    & 44    & 1624  & 18    & 40    & 221.3 & 379.4 & 0.869 \\
    Schoemaker, Paul J.H. & 43    & 35    & 2533  & 17    & 50    & 215.9 & 463.6 & 0.949 \\
    \bottomrule
    \end{tabular}%
  \label{tab:addlabel2}%
\end{table}%

\begin{table}[h]
  \centering
  \caption{Physics researchers data}
    \begin{tabular}{lrrrrrrrr}
    \addlinespace
    \toprule
    Name  & \#pub & \#cited & $T$     & $h$     & $g$     & $j$-index & $j {\mathcal S}$-index & $G_1$ \\
    \midrule
    Alivisatos, A. Paul & 306   & 226   & 43734 & 93    & 209   & 2323.4 & 5009.9 & 0.904 \\
    Wilczek, Frank  & 341   & 269   & 27394 & 82    & 162   & 2028.4 & 4231.8 & 0.850 \\
    Sawatzky, George Albert & 327   & 307   & 21973 & 77    & 139   & 2106.3 & 3921.6 & 0.736 \\
    Jackiw, Roman W. & 210   & 197   & 26478 & 74    & 162   & 1743.9 & 3549.8 & 0.886 \\
    Bradley, Donal D. C. & 426   & 398   & 32292 & 73    & 172   & 2495.5 & 6004.2 & 0.777 \\
    Patel, Popat M. & 894   & 823   & 25560 & 71    & 120   & 3806.8 & 6929.7 & 0.451 \\
    Honscheid, Klaus  & 734   & 682   & 31020 & 66    & 157   & 3182.9 & 7751.0 & 0.670 \\
    Nazarewicz, Witold  & 335   & 306   & 13969 & 66    & 106   & 1712.0 & 3042.9 & 0.655 \\
    Huse, David A. & 172   & 165   & 14096 & 63    & 117   & 1230.6 & 2314.4 & 0.830 \\
    Foxon, C. Thomas & 622   & 540   & 17282 & 62    & 114   & 2337.7 & 4833.1 & 0.603 \\
    Fleming, Robert M. & 181   & 170   & 14600 & 61    & 119   & 1297.3 & 2309.9 & 0.824 \\
    Mättig, Peter  & 675   & 650   & 16750 & 60    & 87    & 2898.0 & 4693.1 & 0.368 \\
    Mikenberg, Giora & 505   & 493   & 14815 & 59    & 84    & 2421.8 & 3750.5 & 0.396 \\
    Minard, Marie-Noelle  & 329   & 314   & 13769 & 59    & 98    & 1809.3 & 3015.4 & 0.572 \\
    Steinhardt, Paul J. & 185   & 174   & 20160 & 58    & 141   & 1358.8 & 2983.0 & 0.894 \\
    Duchovni, Ehud & 444   & 436   & 13006 & 56    & 79    & 2162.2 & 3251.1 & 0.388 \\
    Loebinger, Fred K. & 470   & 466   & 13493 & 53    & 81    & 2252.5 & 3495.4 & 0.387 \\
    Bastard, Gerald  & 222   & 194   & 12926 & 51    & 110   & 1166.8 & 2516.0 & 0.826 \\
    Procaccia, Itamar  & 303   & 281   & 17248 & 48    & 127   & 1449.6 & 3735.0 & 0.807 \\
    Gurtu, Atul  & 373   & 351   & 28061 & 44    & 163   & 1887.7 & 5445.5 & 0.821 \\
    \bottomrule
    \end{tabular}%
  \label{tab:addlabel3}%
\end{table}%


\begin{thebibliography}{}

\bibitem[\protect\citeauthoryear{%
{Bar-Ilan}%
, Levene%
\BCBL{}\ \BBA{} Lin%
}{%
{Bar-Ilan}%
\ \protect\BOthers{.}}{%
{\protect\APACyear{2007}}%
}]{%
BARI07}%
\APACinsertmetastar{%
BARI07}%
{Bar-Ilan}, J.%
, Levene, M.%
\BCBL{}\ \BBA{} Lin, A.%
%
\unskip\
\newblock
\APACrefYearMonthDay{2007}{January}{}.
\newblock
\BBOQ{}\APACrefatitle{Some measures for comparing citation databases}{Some
  measures for comparing citation databases}.\BBCQ{}
\newblock
\APACjournalVolNumPages{Journal of Informetrics}{1}{}{26--34}.
\PrintBackRefs{\CurrentBib}

\bibitem[\protect\citeauthoryear{%
Bartneck%
\ \BBA{} Kokkelmans%
}{%
Bartneck%
\ \BBA{} Kokkelmans%
}{%
{\protect\APACyear{2011}}%
}]{%
BART11}%
\APACinsertmetastar{%
BART11}%
Bartneck, C.%
\BCBT{}\ \BBA{} Kokkelmans, S.%
%
\unskip\
\newblock
\APACrefYearMonthDay{2011}{April}{}.
\newblock
\BBOQ{}\APACrefatitle{Detecting h-index manipulation through self-citation
  analysis}{Detecting h-index manipulation through self-citation
  analysis}.\BBCQ{}
\newblock
\APACjournalVolNumPages{Scientometrics}{87}{}{85--98}.
\PrintBackRefs{\CurrentBib}

\bibitem[\protect\citeauthoryear{%
Bi{H}ui%
, Li{M}ing%
, Rousseau%
\BCBL{}\ \BBA{} Egghe%
}{%
Bi{H}ui%
\ \protect\BOthers{.}}{%
{\protect\APACyear{2007}}%
}]{%
BIHU07}%
\APACinsertmetastar{%
BIHU07}%
Bi{H}ui, J.%
, Li{M}ing, L.%
, Rousseau, R.%
\BCBL{}\ \BBA{} Egghe, L.%
%
\unskip\
\newblock
\APACrefYearMonthDay{2007}{March}{}.
\newblock
\BBOQ{}\APACrefatitle{The {R}- and {AR}-indices: {C}omplementing the
  h-index}{The {R}- and {AR}-indices: {C}omplementing the h-index}.\BBCQ{}
\newblock
\APACjournalVolNumPages{Chinese Science Bulletin}{52}{}{855--863}.
\PrintBackRefs{\CurrentBib}

\bibitem[\protect\citeauthoryear{%
Bornmann%
\ \BBA{} Daniel%
}{%
Bornmann%
\ \BBA{} Daniel%
}{%
{\protect\APACyear{2007}}%
}]{%
BORN07b}%
\APACinsertmetastar{%
BORN07b}%
Bornmann, L.%
\BCBT{}\ \BBA{} Daniel, H.%
%
\unskip\
\newblock
\APACrefYearMonthDay{2007}{}{}.
\newblock
\BBOQ{}\APACrefatitle{What do we know about the {\em h} index?}{What do we know
  about the {\em h} index?}\BBCQ{}
\newblock
\APACjournalVolNumPages{Journal of the American Society for Information Science
  and Technology}{58}{}{1381–1385}.
\PrintBackRefs{\CurrentBib}

\bibitem[\protect\citeauthoryear{%
Bornmann%
, Mutz%
, Hug%
\BCBL{}\ \BBA{} Daniel%
}{%
Bornmann%
\ \protect\BOthers{.}}{%
{\protect\APACyear{2011}}%
}]{%
BORN11}%
\APACinsertmetastar{%
BORN11}%
Bornmann, L.%
, Mutz, R.%
, Hug, S.%
\BCBL{}\ \BBA{} Daniel, H.%
%
\unskip\
\newblock
\APACrefYearMonthDay{2011}{}{}.
\newblock
\BBOQ{}\APACrefatitle{A multilevel meta-analysis of studies reporting
  correlations between the h index and 37 different h index variants}{A
  multilevel meta-analysis of studies reporting correlations between the h
  index and 37 different h index variants}.\BBCQ{}
\newblock
\APACjournalVolNumPages{Journal of Informetrics}{5}{}{346–-359}.
\PrintBackRefs{\CurrentBib}

\bibitem[\protect\citeauthoryear{%
Chatfield%
}{%
Chatfield%
}{%
{\protect\APACyear{1996}}%
}]{%
CHAT96}%
\APACinsertmetastar{%
CHAT96}%
Chatfield, C.%
%
\unskip\
\newblock
\APACrefYear{1996}.
\newblock
\APACrefbtitle{The Analysis of Time Series: An Introduction}{The analysis of
  time series: An introduction}\ (\PrintOrdinal{5th}\ \BEd).
\newblock
\APACaddressPublisher{London}{Chapman \& Hall}.
\PrintBackRefs{\CurrentBib}

\bibitem[\protect\citeauthoryear{%
Diaconis%
\ \BBA{} Graham%
}{%
Diaconis%
\ \BBA{} Graham%
}{%
{\protect\APACyear{1977}}%
}]{%
DIAC77}%
\APACinsertmetastar{%
DIAC77}%
Diaconis, P.%
\BCBT{}\ \BBA{} Graham, R.%
%
\unskip\
\newblock
\APACrefYearMonthDay{1977}{}{}.
\newblock
\BBOQ{}\APACrefatitle{Spearman's footrule as a measure of disarray}{Spearman's
  footrule as a measure of disarray}.\BBCQ{}
\newblock
\APACjournalVolNumPages{Journal of the Royal Statistical Society, Series B
  (Methodological)}{39}{}{262--268}.
\PrintBackRefs{\CurrentBib}

\bibitem[\protect\citeauthoryear{%
Egghe%
}{%
Egghe%
}{%
{\protect\APACyear{2006}}%
}]{%
EGGH06}%
\APACinsertmetastar{%
EGGH06}%
Egghe, L.%
%
\unskip\
\newblock
\APACrefYearMonthDay{2006}{April}{}.
\newblock
\BBOQ{}\APACrefatitle{Theory and practise of the g-index}{Theory and practise
  of the g-index}.\BBCQ{}
\newblock
\APACjournalVolNumPages{Scientometrics}{69}{}{131--152}.
\PrintBackRefs{\CurrentBib}

\bibitem[\protect\citeauthoryear{%
Hirsch%
}{%
Hirsch%
}{%
{\protect\APACyear{2005}}%
}]{%
HIRS05}%
\APACinsertmetastar{%
HIRS05}%
Hirsch, J.%
%
\unskip\
\newblock
\APACrefYearMonthDay{2005}{November}{}.
\newblock
\BBOQ{}\APACrefatitle{An index to quantify an individual's scientific research
  output}{An index to quantify an individual's scientific research
  output}.\BBCQ{}
\newblock
\APACjournalVolNumPages{Proceedings of the National Academy of Sciences of the
  United States of America}{98}{}{16569--16572}.
\PrintBackRefs{\CurrentBib}

\bibitem[\protect\citeauthoryear{%
Motulsky%
}{%
Motulsky%
}{%
{\protect\APACyear{1995}}%
}]{%
MOTU95}%
\APACinsertmetastar{%
MOTU95}%
Motulsky, H.%
%
\unskip\
\newblock
\APACrefYear{1995}.
\newblock
\APACrefbtitle{Intuitive Biostatistics}{Intuitive biostatistics}.
\newblock
\APACaddressPublisher{Oxford}{Oxford University Press}.
\PrintBackRefs{\CurrentBib}

\bibitem[\protect\citeauthoryear{%
Redner%
}{%
Redner%
}{%
{\protect\APACyear{1998}}%
}]{%
REDN98}%
\APACinsertmetastar{%
REDN98}%
Redner, S.%
%
\unskip\
\newblock
\APACrefYearMonthDay{1998}{}{}.
\newblock
\BBOQ{}\APACrefatitle{How popular is your paper? {A}n empirical study of the
  citation distribution}{How popular is your paper? {A}n empirical study of the
  citation distribution}.\BBCQ{}
\newblock
\APACjournalVolNumPages{European Physical Journal B}{4}{}{131--134}.
\PrintBackRefs{\CurrentBib}

\bibitem[\protect\citeauthoryear{%
Segal%
}{%
Segal%
}{%
{\protect\APACyear{2006}}%
}]{%
SEGA06}%
\APACinsertmetastar{%
SEGA06}%
Segal, U.%
%
\unskip\
\newblock
\APACrefYearMonthDay{2006}{July}{}.
\newblock
\BBOQ{}\APACrefatitle{Fair bias}{Fair bias}.\BBCQ{}
\newblock
\APACjournalVolNumPages{Economics and Philosophy}{22}{}{213--229}.
\PrintBackRefs{\CurrentBib}

\end{thebibliography}
\end{document}